# Some properties of the $p_T$ regions observed at the LHC energies.

Mais Suleymanov

COMSATS University Islamabad,
Park Road Islamabad, Pakistan
e-mail: mais_suleymanov@comsats.edu.pk

**Abstract**

The inclusive spectra of the charged particles, $\pi^0$- and $\eta$- mesons produced in the $pp$ collisions at LHC energies were fitted with the exponential functions and analyzed. The data revealed several $p_T$ regions, which could be characterized by the length of the regions $L_K^c$ and two free fitting parameters $a_K^c$ and $b_K^c$. The study of the $L_K^c$ dependences of the parameters $a_K^c$ and $b_K^c$ and energy dependences of the $L_K^c$, $a_K^c$ and $b_K^c$ shown that the regions are divided into two groups. The values of the $L_K^c$ and $b_K^c$ for the first group don't depend on colliding energy and the properties of the particles (though the values of $a_K^c$ increase linearly with energy), however, the second group regions' characteristics show strong dependences. We found that the ratio of the $\eta$- to $\pi^0$- meson lengths is approximately equal to the ratio of their masses: $<L_\eta>:<L_{\pi^0}> \cong m_\eta:m_{\pi^0}$. Assuming that the $L_K^c$ values are directly proportional to the string tension, this result could be the evidence of parton string fragmentation dynamics, since in string theory the masses of elementary particles and their energy are defined by the intensity of string vibration and strangeness of the string stretch. The increasing lengths for the $\eta$-mesons' regions are accompanied with increasing values of $b_k^c$ parameter. Taking into account that $Q^2 \cong \frac{1}{(b_K^c)^2}$ we have concluded that the $\eta$-mesons were produced at lesser values of $\alpha_S$ compare to $\pi^0$-mesons. The results show that the lengths of the first group of regions are ~3-5 times greater than the lengths of neighboring lower $p_T$ regions. The second group's lengths are ~2 times greater than the lengths of neighboring lower $p_T$ region. In the framework of string fragmentation and hadronization dynamics, it could mean that the group *I* is produced through string decay into ~ 3-5 strings, whereas group *II* is produced through decay into ~2 strings. It is possible that the observed two region groups are formed under influence of following factors: energy of the collisions (which defines the tension of the original strings), mass of the produced particles (determined by the vibration of the origin strings), the charge and quantum number of the original strings. The existence of 2 groups of regions could hint to two types of strings responsible for the observed groups.

*1. Introduction*

The $p_T$ distribution of invariant differential yield of the charged primary particles in $pp$ collisions at $\sqrt{s_{NN}} = $ 0.9, 2.76, 7.0 TeV and in Pb-Pb collisions (for six centrality bins) at $\sqrt{s_{NN}} = 2.76$ were disscussed in [1]. The data have been fitted with the exponential functions ($y = ae^{-bp_T}$, here $a$ and $b$ free fitting parameters) and analyzed. They concluded that the data revealed several $p_T$ regions, which reflect the features of fragmentation and hadronization of partons through the string dynamics. That is because the values of $\alpha_S$ (which could be defined by the free fitting parameter $b$ as $\alpha_S \cong \left[ln\left(\frac{q^2}{\Lambda^2}\right)\right]^{-1}$, $\Lambda \cong 0.2$ GeV/c and $Q^2 = -q^2 \cong \frac{1}{b^2}$) depends on $p_T$ and increases during the transition from high $p_T$ region to neighboring low $p_T$ one. The $p_T$ dependence of $\alpha_S$ is typical for the QCD quark string ($\frac{1}{r^2} \sim Q^2 = -q^2$ here $r$ is a distance between quarks in the string[1]). The length of the regions also depends on $p_T$ and the ratio between the lengths of high $p_T$ regions and neighboring low $p_T$ regions show the correspondence $L_{higher}: L_{lower} \cong 3$ or 5 . That is why the length of region was considered a parameter proportional to the string tension.

---
[1] It has been taken that the free fitting parameter $b$ could characterize the distance among the quarks in a string :

It was also concluded in [1] that the most energetic hadrons / partons / strings (with highest tension) are produced in high $p_T$ region ( $p_T > 17 \div 20\ GeV/c$ ) and weakly modified by the medium due to the effect of nuclear transparency. The highest density of the strings is seen in region II ($4 \div 6\ GeV/c < p_T < 17 \div 20\ GeV/c$ ). This high density causes string fusion and a collective phenomenon, which led to the new string formation in the *II* region of $p_T$. The first $p_T$ region ($p_T < 4 \div 6\ GeV/c$) has the maximum number of hadrons and minimum number of strings due to the direct hadronization of low energy strings into two quark systems i.e mesons. It is very interesting to study, how the above-mentioned results will change for the neutral particles, because in paper [1] only charged particles were considered.

In this letter we have applied the same technique as used in [1] to analyses the $p_T$-dependences of invariant cross sections for neutral mesons ($\pi^0$- and $\eta$- mesons) produced in *pp* collisions at LHC energies.

The $\pi^0$-meson is of special interest because it is the lightest hadron and abundantly produces and at LHC energies below 20 GeV/c dominantly originates from gluon fragmentation, above above 20 GeV/c quark fragmentation starts to play a role [2]. Pion production at the RHIC is considered to be dominated by gluon fragmentation only for $p_T < 5$–$8$ GeV/*c* (I $p_T$ region in the paper [1]), at LHC energies it should remain dominant for $p_T < 100$ GeV/*c* [3]. Theoretical estimates suggest that the fraction of pions originating from gluon fragmentation remains above 75% in the $p_T$ range up to 30 GeV/*c* [3].

In the quark model, the $\pi^0$ – meson consists of light-flavor quark-antiquark pairs $u\tilde{u}$ and $d\tilde{d}$, whereas the $\eta$-meson additionally contains hidden strangeness $s\tilde{s}$. So, the comparison of results of charged particles, with the results of $\pi^0$- and $\eta$-mesons could give new information about the properties of $p_T$- regions observed in [1]. The last is very important for the patron string fragmentation dynamics was discussed in the paper [1]. The mesons are quark and anti-quark system and could be considered as a natural string. The study is also important because the string theory defines the masses of elementary particles by vibration energy of the strings, so the massive particle production depends on intensive string vibrations. As mentioned above the length of regions $L_K^c$ is being considered as direct proportional to the tension of strings. So $L_K^c$- and *s*- dependences properties of $p_T$ regions for particles with different characteristics could be used to check some general points of string theory.

We have studied the invariant cross sections for neutral particles $\pi^0$- and $\eta$- mesons as a function of $p_T$ produced in *pp* collisions. The data at 0.9 TeV and 7 TeV is used from [3], for 2.76 TeV from [2] and for 8 TeV from [4]. Tree variables: length of the region $L_K^c$ and two free fitting parameters $a_K^c$ and $b_K^c$ (the upper index *c* shows the type of events and the lower index *K* is for the regions number) were used to describe the regions. The values of the $L_K^c$, $a_K^c$ and $b_K^c$ were found by fitting the distributions with a simple function $y = a_K^c e^{b_K^c p_T}$. To get the best fitting results we varied the values of the $p_T$ from $p_T^{min}$ to $p_T^{max}$ for separate regions with the original distribution. The values of the $p_T^{min}$ and $p_T^{max}$ were used to define the boudries[2] of regions $J_K^c$ and to extract the values of the $L_K^c$. The best results of the fitting are taken to analyses the properties of the $p_T$ regions.

## *2 . Data for the charged particles , $\pi^0$- , and $\eta$- mesons produced in the pp collisions at 0.9 TeV.*

---

$\frac{1}{r^2} \sim Q^2 = -q^2 \cong \frac{1}{b^2} \cong <p_T>^2$

[2] The values of $J_K^c$ were calculated using experimentally mesured values of $p_T^{min}$ (the value of $p_T^{min}(I)$ for the *I* regions was taken as $p_T^{min}(I) = 0$) and $p_T^{max}$ as $J_{I-II}^c = (p_T^{max}(I) + p_T^{min}(II))/2$ for the region *I* and *II* , $J_{II-III}^c = (p_T^{max}(II) + p_T^{min}(III))/2$ for the region *II* and *III* and so on. Then by dint of the values of $J_K^c$ the values of the lengths for the different $p_T$ regions $L_K^c$ ( *K* is the number of regions) in various collisions *c* were defined. The values of $L_K^c$ were calculated as $L_I^c = J_{I-II}^c$ ; $L_{II}^c = J_{II-III}^c - J_{I-II}^c$ and $L_{III}^c = J_{III-IV}^c - J_{II-III}^c$ and so on.

The letter [3] presents the first measurement of neutral meson production in proton-proton collisions at center-of mass energies of $\sqrt{s}=0.9$ TeV and 7 TeV in a wide $p_T$ range with the ALICE detector. The Fig. 1 shows the $p_T$ distributions for the charged particles[3] and neutral pions[4] produced in the $pp$ collisions at 0.9 TeV. As we have mentioned above the distributions were fitted with a simple function $y = a_K^c e^{b_K^c p_T}$, where $a_K^c$ and $b_K^c$ are free fitting parametes (the index $K$ shows the number of region and $c$ indicates the type of events in which the regions were formed)[5]. To get the best fitting results we varied the values of the $p_T$ from $p_T^{min}$ to $p_T^{max}$ and fitted separate regions with the original distribution. The best results of the fitting are shown in the Table 1. The data for the charged particles (were taken from the paper [1]) were shown using the index $c=3$. The data for the neutral pions indicated by index $c=31$. The boundary values of the $p_T$ for the extracted regions $J_K^c$ are also shown by vertical lines in the figure. The dashed lines are used for the charged particles ($c=3$), and solid line is used for the neutral mesons ($c=31$).

      From the data presented in the Table 1, one can see that the distributions contain several cirten $p_T$ regions. It is possible to define only the characteristics for the two distinct $p_T$ regions of the charged particles and only for one $p_T$ region for the neutral pions (events with $c=31$). In the paper [1] it was discussed that for charged particles impossible to define the characteristics for the *III* region due to mesument limit, the experimental mesument allowed to see charged particles $p_T$ distribution in the $p_T$ interval of : 05÷31.2 TeV/c [3]. But for the $\pi^0$-meson produced in $pp$ collisions at energy 0.9 TeV we have experimentally mesured $p_T$ distribution in the shorter $p_T$ interval of : 0.5÷5.4 GeV/c [3].

      In the Table 1 the simbol $J_K^c$ shows the boundary values of the regions (see footnote 3). The values of the $J_K^c$ were used to calculate the values of the $L_K^c$ (see footnote 3). The values of the $\chi^2/ndf$ and Prob. (probability) are also included in the Table 1 tagather with the values of free fitting parameters $a_K^c$ and $b_K^c$.

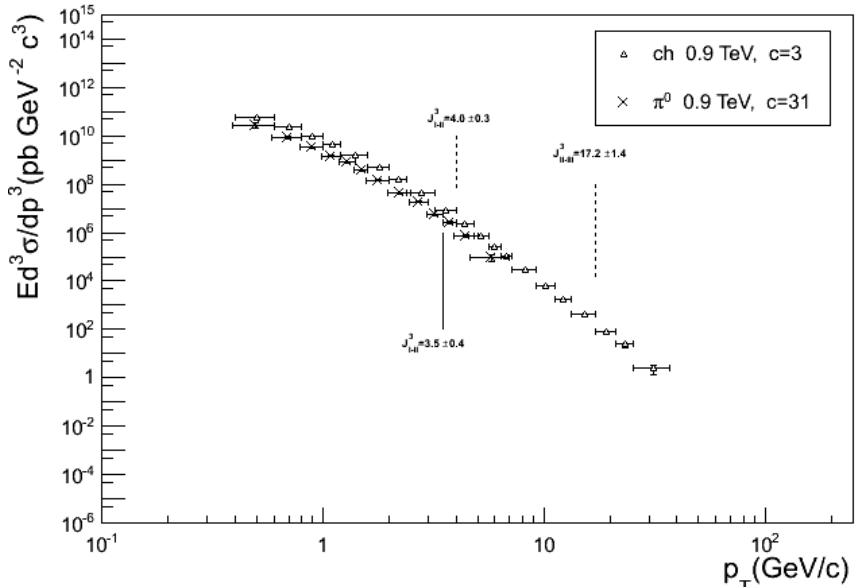

Fig. 1 The $p_T$ distributions for the charged particles and neutral pions produced in the $pp$ collisions at 0.9 TeV.

---

[3] The experimental data were taken from the HEP Data: https://hepdata.net/record/ins896764 . The distributions for the charged particles (c=3) were multiplied to $\sigma_{inel}$=5.3 $10^{10}$ pb to be able to compare with the one for the pions (c=31). The value of $\sigma_{inel}$ have been got from the figure 1 in the paper [5].
[4] The data were taken from the HEP Data: https://hepdata.net/record/ins1116147
[5] To fit the distributions we used ROOT soft (version 5.34/02 , 21$^{st}$ September 2012), taking into account the statistical and systematic uncertainties of the measurement.

Table 1. The best fitting results for *pp* collisions at 0.9 TeV.

| $c$ ↓ | $K \rightarrow$ | I | II | III |
|---|---|---|---|---|
| $c=3$[6] | $p_T^{min} \div p_T^{max}$<br>$J_K^c$<br>$L_K^c$<br>$\chi^2/ndf;Prob.$<br>$a_K^c (pb\ GeV^2\ c^3)$<br>$b_K^c\ (GeV/c)^{-1}$ | (0.5±0.1) ÷ (3.6±0.4)<br>$J_{I,II}$= 4.0±0.3<br>4.0±0.3<br>5.972/7;0.5431<br>(2.0±0.4)10$^{11}$<br>3.0±0.1 | (4.4±0.4)÷(15.2±2.0)<br>$J_{II,III}$= 17.2±1.4<br>13.2±1.4<br>5.805/6;0.4454<br>(7.6±2.7)10$^7$<br>0.87 ± 0.05 | (19.2±2.0) ÷ (31.2±6.0)<br>$J_{III,IV}$= 31.2±6.0<br>14.0±6.2<br>0.0005462/1; 0.9814<br>(2.2±6.8) 10$^4$<br>0.29±0.14 |
| $c=31$ | $p_T^{min} \div p_T^{max}$<br>$J_K^c$<br>$L_K^c$<br>$\chi^2/ndf;Prob.$<br>$a_K^c(pb\ GeV^2\ c^3)$<br>$b_K^c\ (GeV/c)^{-1}$ | (0.5±0.1) ÷(3.2±0.3)<br>$J_{I,II}$= (3.5±0.4)<br>(3.5±0.4)<br>8.451/8;0.3907<br>(6.6±2.1)10$^{10}$<br>(3.1±0.2) | (3.7±0.3)÷(5.4±1.2)<br>-<br>-<br>0.006141/1;0.9375<br>(1.1±3.4)10$^9$<br>1.6±0.8 | -<br>-<br>-<br>-<br>-<br>- |

The data from the Fig. 2 and Fig. 3 show that the values of the free fitting parameters $a_K^c$ and $b_K^c$ decrease exponentially with the lengths of the regions (the lines has drown by hand). The results for the energy dependences will be discussed in the chapter 6.

### 3. Data for the charged particles, $\pi^0$- and $\eta$- mesons produced in the pp collisions at 2.76 TeV.

The Fig. 4 shows the invariant cross sections for the $\pi^0$- and $\eta$- mesons (index *c*=11 and 12 respectively) production in *pp* collisions at 2.76 TeV [5][7]. To compare the data, the same ones[8] for the charged particles produced in the *pp* collisions at 2.76 TeV (index *c*=1 as in paper [1]) are also drown in the figure[9].

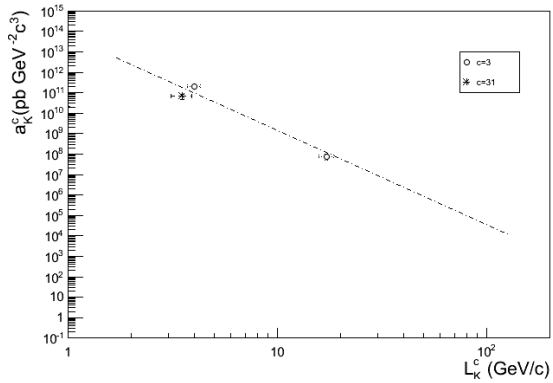
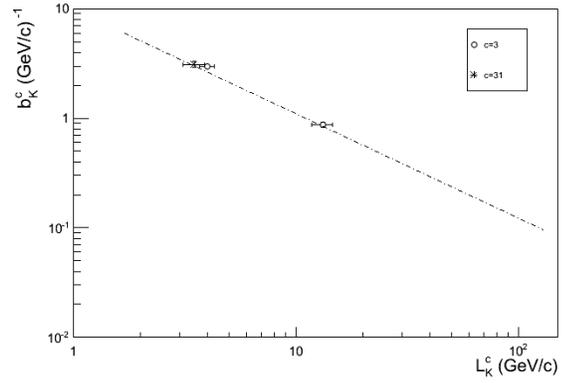

Fig. 2                                                                                     Fig.3

---

[6] In the paper [1] the distributions of the yield for the charged particles were fitted. That is why the values of the parameters $a_K^3$ have to multiple to the inelastic cross section $\sigma_{inel}$=5.3 10$^{10}$ pb to be able to compare with the data for the parameter $a_K^{31}$. The values of the $\sigma_{inel}$ was taken from the figure 1 in the paper [5].
[7] The data were got from the HEP Data: https://hepdata.net/record/ins1512110
[8] The experimental data were taken from the HEP Data: https://hepdata.net/record/ins1088823
[9] To get the distribution the data on yield (which were taken from the HEP Data: https://hepdata.net/record/ins1088823) for the charged particles produced in *pp* collisions at 2.76 TeV (from the paper [1]) had been multiplied to the cross section of inelastic charged particle production ($\sigma_{ind}$) in the collisions. The value of $\sigma_{inel}$ =60 mb have been taken from the figure 1 in the paper [5].

It seems visually that there are not any essential differences between the distributions. Though the results of the fitting these distributions pointed out to existing some differences [10]. The Table 2 shows the best fitting results for these distributions. To compare the data, the best fitting results of the ones for the charged particles produced in *pp* collisions at 2.76 TeV [1] are also included to the table. The Table 2 says us that the distribution for the $\pi^0$-meson (*c*=11) contains 3 different regions as well as the distribution for the charged particles (*c*=1, see Table 2)[11] though the data for the $\eta$-mesons (*c*=12) continues 2 regions. But the data for the neutral mesons allow to define the characteristics for 2 regions only. The values for $p_T$ at boundaries of the regions $J_K^c$ for the: *I* and *II* is $J_{I-II}^c$; *II* and *III* does $J_{II-III}^c$; *III* and *IV* does $J_{III-IV}^c$ (upper indexes *c* show the type of event in which the region was formed and down ones *K* do the number of the region) also shown by vertical lines in the figure together with the values of $J_K^c$. The dashed-dot lines are used for the charged particles (*c*=1), dashed lines do for the $\pi^0$-mesons (*c*=11) and solid lines are used for the $\eta$-mesons (*c*=12). The values of the $J_K^c$ were used to calculate the ones for the $L_K^c$ as $L_I^c = J_{I-II}^c$; $L_{II}^c = J_{II-III}^c - J_{I-II}^c$ and $L_{III}^c = J_{III-IV}^c - J_{II-III}^c$ and so on (see Table 2).

The data from the Fig.5 and Fig. 6[12] show that the values of the free fitting parameter $b_K^c$ decrease exponentially by length ($L_K^c$) having behavior almost like to one for the charged particles. The behavior of the $L_K^c$ dependence for the parameter $a_K^c$ shows some deviations from the behavior for the charged particles of the events with *c*=12 (eta mesons) in the first region and in the event with *c*=11 (neutral pions) from the *II* region. These deviations connected with the increasing the length (*I* region) for the $\eta$-mesons and decreasing the $L_{II}^{11}$ for the neutral pions by comparison with charged particles. These changing of the values of $L_K^c$ do not affect the values of $b_K^c$. The *s*-dependences for the $L_K^c$, $a_K^c$ and $b_K^c$ will be discussed in the capture 6.

### 4. Data for the charged particles, $\pi^0$- and $\eta$- mesons produced in the pp collisions at 7 TeV.

The Fig. 7 shows the first measurements of the invariant differential cross sections of inclusive $\pi^0$ (*c*=41) and $\eta$ meson (*c*=42) production at mid-rapidity in proton–proton collisions at $\sqrt{s}$ = 7 TeV [3][13]. The distribution for the charged particles (*c*=4) from the *pp* collisions at $\sqrt{s}$ = 7 TeV has been drown in the figure too [6][14,15]. The distributions were fitted by the exponential function $y = a_K^c e^{-b_K^c p_T}$ ($a_K^c$ and $b_K^c$ the free fitting parameters). To get the best fitting results we varied the values of the $p_T$ from $p_T^{min}$ to $p_T^{max}$ (as in the paper [1]) and fitted separate regions with the original distribution. The best results of the fitting are shown in the Table 3. From the data presented in the table one can see that the fit extract several distinct regions of $p_T$ distributions. The values for $p_T$ at boundaries of the regions $J_K^c$ [16] for the: *I* and *II* is $J_{I-II}^c$; *II* and *III* does $J_{II-III}^c$; *III* and *IV* is $J_{III-IV}^c$ (upper indexes *c* show the type of event in which the region was formed and down ones *K* do the number of the region) also shown by vertical lines in the figure together with the values of the $J_K^c$. The dashed-dot lines are used for the charged particles (*c*=4), dashed lines do for the $\pi^0$-mesons (*c*=41) and solid lines are used for the $\eta$-mesons (*c*=42).

The data from the Fig. 8 and Fig. 9 show that the values of the free fitting parameters $a_K^c$ and $b_K^c$ decrease exponentially with the lengths of the regions (the lines has drowned by hand). The energy dependences of the $L_K^c$, $a_K^c$ and $b_K^c$ will be discussed in the chapter 6.

---

[10] As we have mentioned above the distributions were fitted by the exponential function ($y = a_K^c e^{b_K^c}$) in the intervals of $p_T$ between: $p_T^{min} \div p_T^{max}$, here the $p_T^{min}$ and $p_T^{max}$ are the minimum and the maximum values of experimentally mesured $p_T$, obtaned as a result of variation of $p_T$ values to get the best fitting results.

[11] It is necessary to mention that the data for charged particles produced in the *pp* collisions at 2.76 TeV were measured in the $p_T$ interval:0.525 ÷99.3 GeV/c though the data for $\pi^0$- and $\eta$- mesons were done in the intervals of 0.5 ÷ 37; 0.7-18 respectively.

[12] The second point for the eta mesons could be defined with errors great than 100%.

[13] The experimental data were taken from the: https://hepdata.net/record/ins1116147

[14] The experimental data were taken from the HEP Data: https://hepdata.net/record/ins896764

[15] In the paper [6] the experimental data on $p_T$ distributions of the invariant charged particles differential yield averaged over the pseudorapidity |$\eta$| < 1 for charged primary particles produced in *pp* collisions (index c=1) at 7 TeV were presented. The data were multiplied to the inelastic cross sections of the *pp* collisions at 7 TeV, $\sigma_{inel}$ =72 mb (was taken from the figure 1 in the paper [5]) to compare with the data for the neutral mesons.

[16] The values of the $J_K^c$ were used to calculate the ones for the $L_K^c$ as $L_I^c = J_{I-II}^c$; $L_{II}^c = J_{II-III}^c - J_{I-II}^c$ and $L_{III}^c = J_{III-IV}^c - J_{II-III}^c$ and so on.

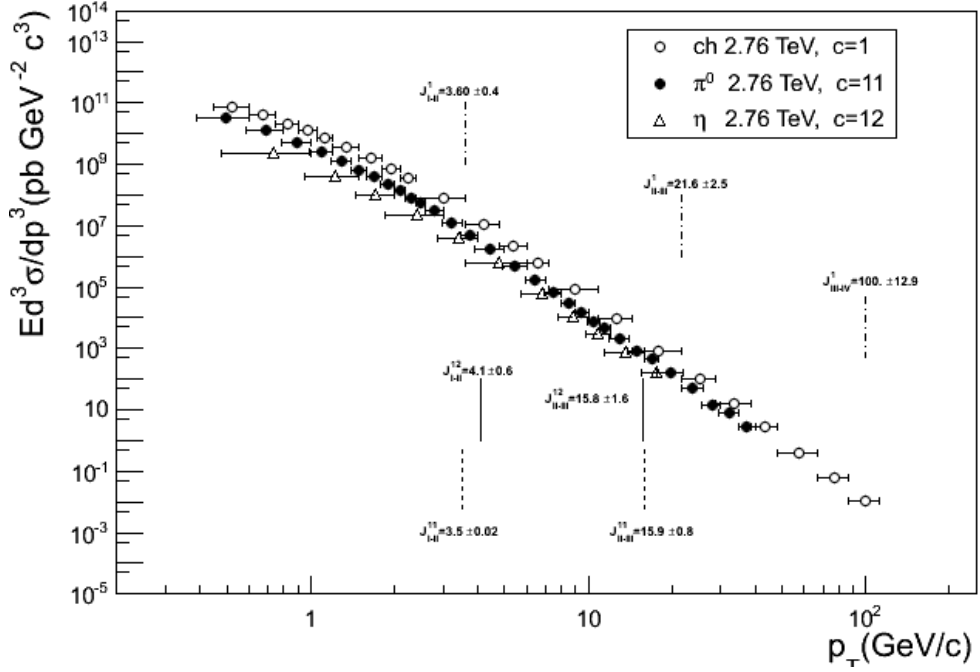

Fig. 4 The invariant cross sections for the $\pi^0$- and $\eta$- mesons ( index $c$=11 and 12 respectively) production in $pp$ collisions at 2.76 TeV

### 5. Data for the charged particles, $\pi^0$- and $\eta$- mesons produced in the pp collisions at 8 TeV.

The Fig. 10 shows the invariant cross sections for neutral mesons ($\pi^0$- and $\eta$- mesons, index $c$=51 and 52 respectively)[17] [4] and charged particles (index $c$=5)[18] (with $|\eta| < 2.5$) production in $pp$ collisions at 8 TeV [7] (in the events with a number of charged particle $n_{ch} \geq 1$). One can see that visually the differences between the distributions are mainly in the interval of $p_T < 2.0$ GeV/c. Though the results of fitting these distributions pointed out to more differences[19]. The Table 4 shows the best fitting results for the distributions from the Fig.10. There exist several $p_T$ regions with spesial properties. The regions will be characterized using 3 variables: length of the region $L_K^c$ and the values of two free fitting parameters $a_K^c$ and $b_K^c$. To get the values of the $L_K^c$ the ones of the $p_T$ at boundaries of the regions $J_K^c$ ( $J_{I-II}^c$ (for I and II regions), $J_{II-III}^c$ (for II and III ones), $J_{III-IV}^c$ (for III and IV regions) and $J_{IV-V}^c$ (for IV and V ones), upper indexes are the values of events $c$ and down ones do the index $K$ - the number of the region) were used. The values of the $J_K^c$ are also shown by vertical lines in the figure. The solid lines are used for the charged particles, dashed lines do for the $\pi^0$-mesons and dashed-dot lines are used for the $\eta$-mesons. The values of the $J_K^c$ were used to calculate the ones for the $L_K^c$ as $L_I^c = J_{I-II}^c$ ; $L_{II}^c = J_{II-III}^c - J_{I-II}^c$ and $L_{III}^c = J_{III-IV}^c - J_{II-III}^c$ and so on (see Table 4).

---

[17] The data were got from the HEP Data: https://hepdata.net/record/ins1620477
[18] The data were got from the HEP Data: https: https://hepdata.net/record/ins1426695
[19] The distributions were fitted by the exponential function ($y = a_K^c e^{-b_K^c p_T}$) in the intervals of $p_T$ between: $p_T^{min} \div p_T^{max}$ (here the $p_T^{min}$ and $p_T^{max}$ are the minimum and the maximum values of experimentally mesured $p_T$, obtaned as a result of variation of $p_T$ values to get best fitting results).

Table 2. The best fitting results for *pp* collisions at 2.76 TeV.

| | | | | |
|---|---|---|---|---|
| $c=1$[20] | $p_T^{min} \div p_T^{max}$ | (0.53±0.08)÷(3.0±0.6) | (4.2±0.6)÷(18.0±3.6) | (25.2±3.6)÷(99.3±12.9) |
| | $J_K^c$ | $J_{I,II}$= (3.6±0.4) | $J_{II,III}$= (21.6±2.5) | $J_{III,IV}$= (100.0±12.9) |
| | $L_K^c$ | (3.6±0.4) | (18.0±2.5) | (78.4±13.3) |
| | $\chi^2/ndf$; Prob. | 4.132/8;0.845 | 3.458/4; 0.4842 | 2.47/4;0.65 |
| | $a_K^c$ (pb GeV$^2c^3$) | (3.1±0.6)10$^{11}$ | (1.8±0.8)10$^8$ | (2.4±1.4) 10$^6$ |
| | $b_K^c$ (GeV/c)$^{-1}$ | 3.1±0.2 | 0.74±0.06 | 0.13±0.01 |
| $c=11$ | $p_T^{min} \div p_T^{max}$ | (0.5±0.1) ÷(3.2±0.3) | (3.7±0.3)÷(14.9±1.1) | (16.9±1.1)÷(37.3±2.7) |
| | $J_K^c$ | $J_{I,II}$= (3.5±0.2) | $J_{II,III}$= (15.9±0.8) | $J_{III,IV}$= (-) |
| | $L_K^c$ | (3.5±0.2) | (12.5±0.8) | (-) |
| | $\chi^2/ndf$; Prob. | 10.77/11;0.4631 | 8.00318/8; 0.4332 | 1.126/4;0.8901 |
| | $a_K^c$ (pb GeV$^2c^3$) | (9.5±1.9)10$^{10}$ | (3.9±1.4)10$^7$ | (3.2±2.0) 10$^4$ |
| | $b_K^c$ (GeV/c)$^{-1}$ | 3.0±0.1 | 0.78±0.04 | 0.26±0.03 |
| $c=12$ | $p_T^{min} \div p_T^{max}$ | (0.7±0.3) ÷(3.4±0.6) | (4.8±1.1)÷(13.7±2.3) | (17.8±2.2)÷() |
| | $J_K^c$ | $J_{I,II}$= (4.1±0.6) | $J_{II,III}$= (15.8±1.6) | $J_{III,IV}$= - |
| | $L_K^c$ | (4.1±0.6) | (11.7±1.7) | - |
| | $\chi^2/ndf$; Prob. | 0.9034/3;0.8246 | 0.6984/3; 0.8736 | - |
| | $a_K^c$ (pb GeV$^2c^3$) | (1.1±0.8)10$^9$ | (2.2±2.5)10$^7$ | - |
| | $b_K^c$ (GeV/c)$^{-1}$ | 2.4±0.4 | 0.79±0.11 | - |

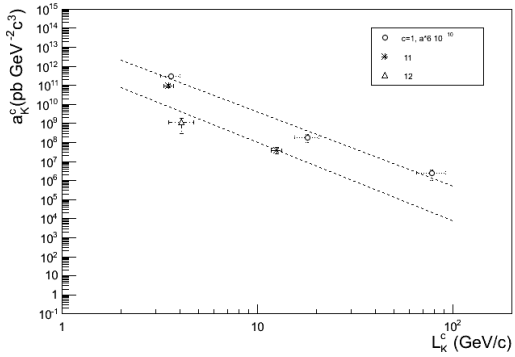
Fig.5
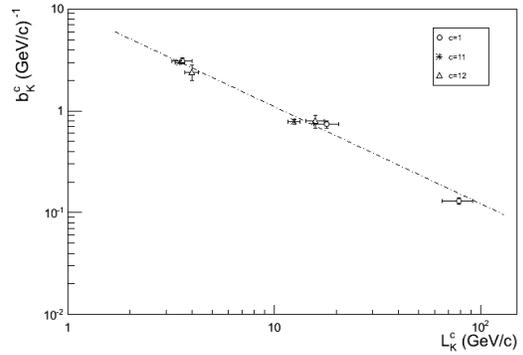
Fig.6

The data from the Table 4 says us that the distribution for the $\pi^0$-mesons contains at least 5 different regions but the distribution for the $\eta$-mesons has 3 regions though for charged particles there are 3 regions (see Table 4). The data from the table allow to define the characteristics for 4 regions in case of $\pi^0$-mesons and for 2 ones in case of $\eta$-mesons[21] and the charged particles.

The Fig. 11 and 12 show behavior of the free fitting parameters $a_K^c$ and $b_K^c$ for the $\pi^0$- and $\eta$-mesons as a function of $L_K^c$. The values of the parameters $a_K^c$ and $b_K^c$ for the charged particles decrease almost exponentially by $L_K^c$ ( the dashed lines in the figure are drown by hand). But the values of the free parameters for $\pi^0$- and $\eta$-mesons as a function of regions' lengths deviate from the exponential. The reason of the deviations is the changing in the lengths for the neutral mesons. The lengths increase for the eta mesons and decrease for the $\pi^0$ -mesons. The energy dependences for the $L_K^c$, $a_K^c$ and $b_K^c$ will be discussed in the next capture.

---
[20] The values of the parameter $a_K^c$ were multiplied to the cross section of inelastic charged particle production ($\sigma_{inel}$) in the collisions, the values of the $\sigma_{inel}$ =60 mb had been taken from Fig. 1 in the paper [5].
[21] To define the characteristics for the fifth region in case of $\pi^0$–mesons and third one for the $\eta$-mesons the new high $p_T$ measurements are required in the interval of $p_T$ > 32.3 GeV/c. Same situation was for the IV region of charged particles produced in the *pp* collisions at 7 TeV [1], to define the characteristics for the IV regions the new high $p_T$ measurements in the interval of $p_T$ > 181 GeV/c were required.

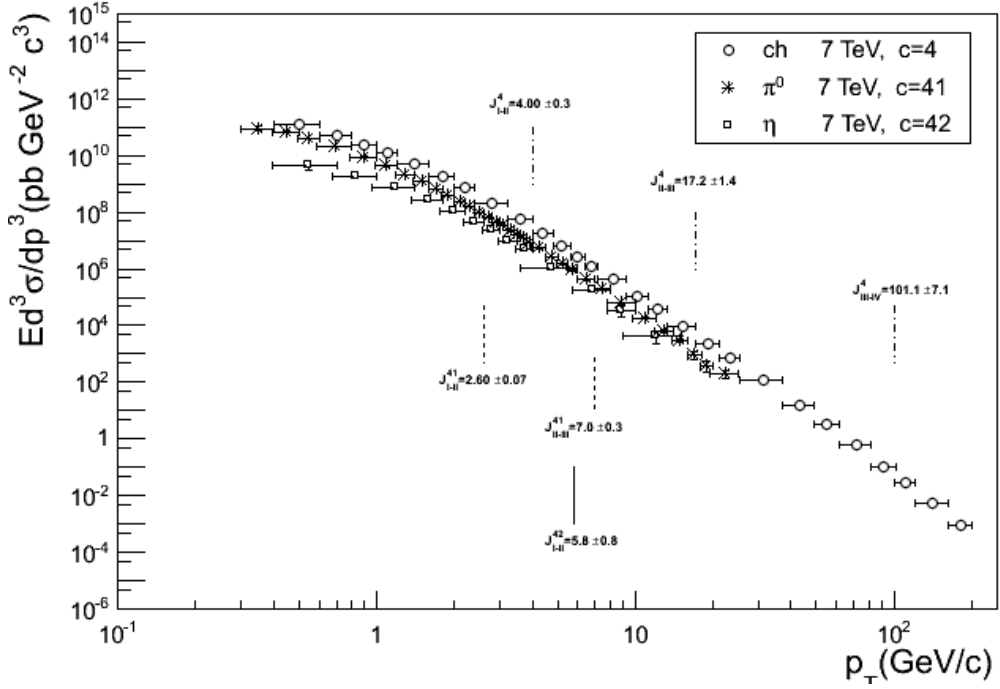

Fig.7 The invariant differential cross sections of inclusive $\pi^0$ ($c$=41) and $\eta$ meson ($c$=42) production at mid-rapidity in proton–proton collisions at $\sqrt{s}$ = 7 TeV.

Table 3. The best fitting results for pp collisions at 7 TeV ($c$=4).

| $c$ ↓ | $K$ → | I | II | III |
|---|---|---|---|---|
| $c=4^{22}$ | $p_T^{min} \div p_T^{max}$ $J_K^c$ $L_K^c$ $\chi^2/ndf$;Prob. $a_K^c$ (pb GeV$^2c^3$) $b_K^c$ (GeV/c)$^{-1}$ | (0.5±0.1)÷(3.6±0.4) $J_{I,II}$= (4.0±0.3) (4.0±0.3) 6.55/7;0.4771 (3.5±0.7)10$^{11}$ 2.7±0.1 | (4.4±0.4)-(15.2±2.0) $J_{II,III}$= (17.2±1.4) (13.2±1.4) 5.632/6;0.4656 (3.6±1.4)10$^8$ 0.76±0.04 | (19.2±2.0)÷(91.2±10.0) $J_{III,IV}$= (101.1±7.1) (83.9±7.2) 4.238/5;0.5157 (3.2±1.1) 10$^4$ 0.146±0.009 |
| $c=41$ | $p_T^{min} \div p_T^{max}$ $J_K^c$ $L_K^c$ $\chi^2/ndf$;Prob. $a_K^c$ (pb GeV$^2c^3$) $b_K^c$ (GeV/c)$^{-1}$ | (0.35±0.05)÷(2.5±0.1) $J_{I,II}$= (2.60±0.07) (2.60±0.07) 9.184/11; 0.6049 (2.5±0.3)10$^{11}$ (3.3±0.1) | (2.7±0.1)÷(6.5±0.5) $J_{II,III}$= (7.0±0.3) (4.4±0.4) 5.552/10;0.8514 (3.0±0.7)10$^9$ (1.47±0.06) | (7.5±0.5)÷(22.2±2.8) - - 2.448/6;0.8743 (9.7±4.4)10$^6$ (0.53±0.04) |

---

[22] The values of parameter $a_K^c$ were multiplied to inelastic cross-section 72 mb was taken from the figure 1 in the paper [5].

| | | | | | |
|---|---|---|---|---|---|
| c=42 | $p_T^{min} \div p_T^{max}$ | (0.5±0.1)÷(4.8±1.2) | (6.8±1.2)÷(12.0±3.0) | - | - |
| | $J_K^c$ | (5.8±0.8) | - | - | - |
| | $L_K^c$ | (5.8±0.8) | - | - | - |
| | $\chi^2/ndf$;Prob. | 2.948/8; 0.9376 | 0.005293/1;0.942 | - | - |
| | $a_K^c (pb\ GeV^2 c^3)$ | $(1.0\pm 0.3)10^{10}$ | $(2.4\pm 7.5)10^7$ | - | - |
| | $b_K^c\ (GeV/c)^{-1}$ | (2.1±0.1) | 0.7±0.4 | - | - |

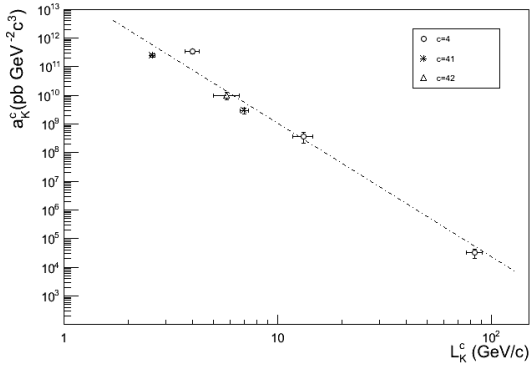

Fig. 8

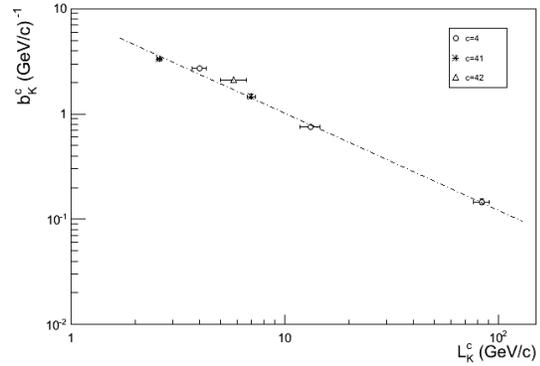

Fig. 9

## 6. Discussion the results.

Let us start from the data presented in the Figs. 2-3; 5-6; 8-9 and 11-12 where the $L_K^c$ – dependences for the free fitting parameters $a_K^c$ and $b_K^c$ are demonstrated. At energies 0.9 TeV the values of the $a_K^c$ and $b_K^c$ decrease exponentially with $L_K^c$ both for charged particles and the neutral mesons. With increasing the energy there are some deviations from the exponentially deceasing behaviors and it is seen that these deviations are connected mainly with changing the lengths for the neutral mesons: for the neutral pions' lengths these parameters decrease; and increase for the eta mesons' lengths as compared with the lengths of charged particles. This result is in favor of the string fragmentation and hadronization dynamics and it will be confirmed by other evidences too, as presented below. The considered neutral mesons are the systems with two quarks only $q\tilde{q}$. The system could be considered as a natural string though the responsible strings to produce the charged particles have to be more complex quark systems. The last could be a reason for the differences (the deviations) between the characteristics of charged particles and neutral mesons.

To discuss the results obtained in the chapters 2-5, the Figs 13-15 are drawn and Tables 5-7 are created. The Figs. 13-15 show the energy dependencies ($s$) for the $p_T$ regions' lengths ($L_K^c$) and for the two free fitting parameters $a_K^c$ and $b_K^c$ respectively in the different events ( the upper index $c$ and the down index $K$ show the various $p_T$ regions ). The lines in the figure are drawn manually. One can see that:

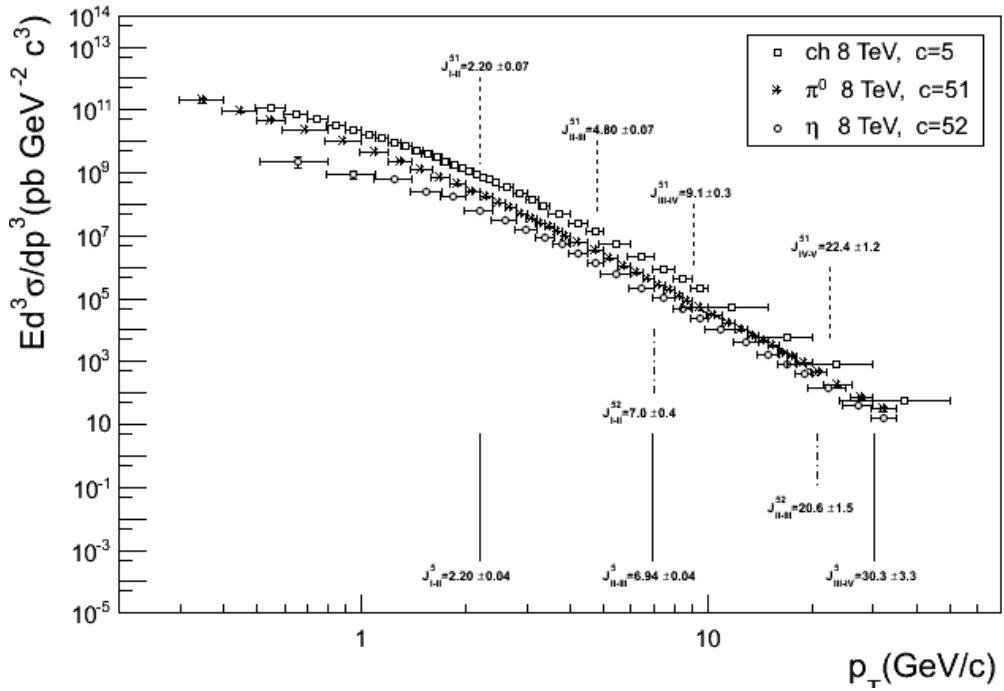

Fig. 10 The invariant cross sections for neutral meson ($\pi^0$- and $\eta$- mesons, index $c=51$ and 52 respectively) and charged particles (index $c=5$) (with $|\eta| < 2.5$) produced in $pp$ collisions at 8 TeV ( in the events with a number of charged particle $n_{ch} \geq 1$).

Table 4. The best fitting results for $pp$ collisions at 8 TeV ($c=5$).

| $c$ ↓ | Region → | I | II | III | IV | V |
|---|---|---|---|---|---|---|
| c=5 | $p_T^{min} \div p_T^{max}$ $J_K^c$ $L_K^c$ $\chi^2/ndf;Prob.$ $a_K^c (pb\ GeV^2c^3)$ $b_K^c\ (GeV/c)^{-1}$ | (0.55±0.05)÷(2.15±0.05) $J_{I,II}$=2.20 ± 0.04 (2.20 ±0.04) 14.73/15;0.4707 (4.7±0.5)10$^{11}$ 3.00±0.07 | (2.25±0.05)÷(6.44± 0.06) $J_{II,III}$= 6.94± 0.04 (4.74 ±0.06) 6.541/10 ; 0.768 (2.5 ±0.4)10$^{10}$ 1.59±0.06 | (7.44±0.06) - - 6.051/5;0.313 (1.2±0.3)10$^7$ 0.38±0.02 | - - - - - - | - - - - - - |
| c=51 | $p_T^{min} \div p_T^{max}$ $J_K^c$ $L_K^c$ $\chi^2/ndf;Prob.$ $a_K^c (pb\ GeV^2c^3)$ $b_K^c\ (GeV/c)^{-1}$ | (0.35 ±0.05 – 2.1±0.1) $J_{I,II}$=2.20±0.07 2.20±0.07 9.148/9; 0.4237 (4.5±0.9)10$^{11}$ 3.7±0.2 | (2.3 ±0.1 – 4.4±0.1) $J_{II,III}$=4.80±0.07 2.6±0.1 2.152/9;0.9888 (7.7±2.2)10$^9$ 1.71±0.09 | (5.2±0.1– 8.7±0.1) $J_{III,IV}$=9.1±0.3 4.3±0.3 0.6687/6;0.9951 (1.7±0.8)10$^7$ 0.88±0.07 | (9.5±0.5- 21.0±1.0) $J_{IV,V}$=22.4±1.2 13.3±1.2 4.46/9;0.8786 (2.5±0.8)10$^6$ 0.43±0.02 | (23.8±2.2-32.3±2.7) - - 0.001818/1 ; 0.996 (2.2±5.2)10$^4$ 0.20±0.08 |
| c=52 | $p_T^{min} \div p_T^{max}$<br>$J_K^c$<br>$L_K^c$<br>$\chi^2/ndf; Prob.$<br>$a_K^c (pb\ GeV^2c^3)$<br>$b_K^c (GeV/c)^{-1}$ | (0.7±0.1 – 6.5±0.5)<br>$J_{I,II}$=7.0±0.4<br>7.0±0.4<br>10.73/12;0.552<br>(3.7±0.7)10⁹<br>1.65±0.06 | (7.5±0.5 -18.9±1.1)<br>20.6±1.5<br>$J_{II,III}$=13.6±1.6<br>3.113/6;0.7945<br>(3.5±1.2)10⁶<br>0.49±0.03 | (22.2±2.8– 32.3±2.7)<br>11.7±3.1<br>-<br>0.05913/1;0.8079<br>(2.1±4.7)10⁴<br>0.22±0.08 | -<br>-<br>-<br>-<br>-<br>- | -<br>-<br>-<br>-<br>-<br>- |

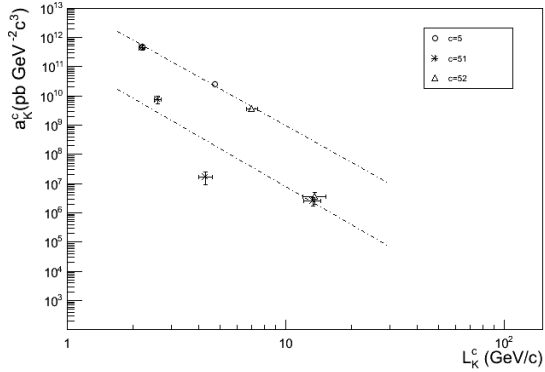   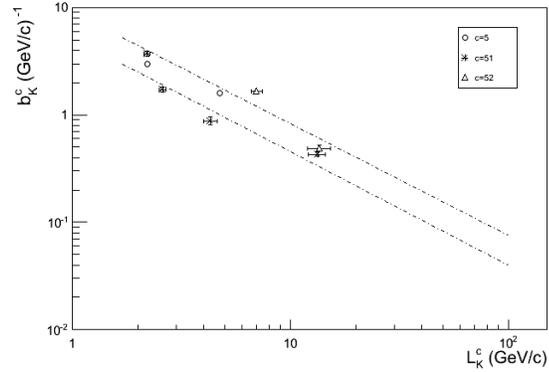

Fig. 11                                               Fig. 12

1. At the energies 0.9 TeV , 2.76 TeV all the points and at the energy 7 TeV the points for the charged particles , corresponding to the values of $L_K^c$, $a_K^c$ and $b_K^c$ lie well on the lines (1)-(3) , points of region *I* are on the line (1), the region *II* on the line (2) and the points of region *III* on the line (3) – "normal" location of the points. There is no any essential dependences on the colliding energy and the characteristics of the produced particles for the values of the $L_K^c$ and $b_K^c$ but there is some liner *s*-dependences for the values of $a_K^c$ in this group of regions - first group of regions.
2. Some deviations (as shown by circles around points, in the figure) from the "normal" location start to appear at energies: 7 TeV for the points corresponding to the values of $L_K^c$, $a_K^c$ and $b_K^c$ of $\eta$-mesons' regions and for the ones for $\pi^0$ –mesons; 8 TeV for all points. These regions will be mentioned as the second group of regions and there are some dependences on the colliding energy and the characteristics of the produced particles for this group of regions.

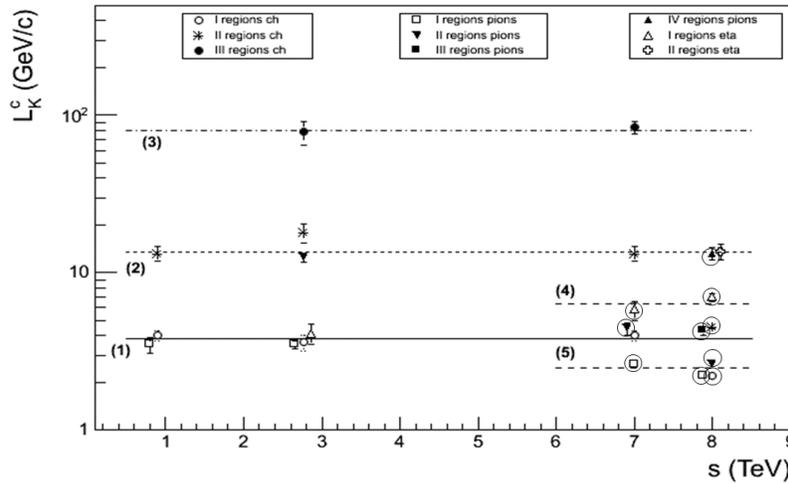

Fig. 13 The *s*-dependences for the values of the $L_K^c$.

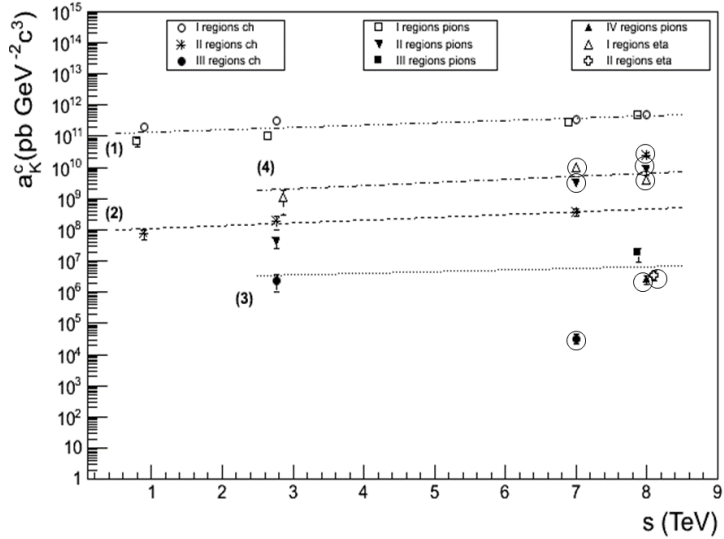

Fig. 14 The $s$-dependences for the values of the $a_K^c$.

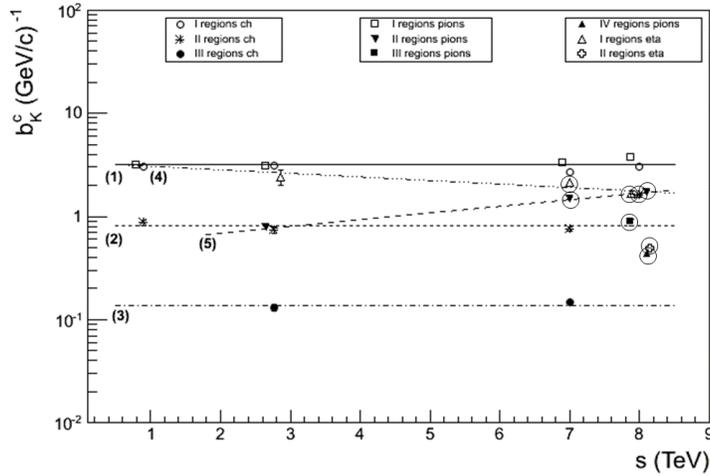

Fig. 15 The $s$-dependences for the values of the $b_K^c$.

3. The lengths of the $p_T$ regions decrease with energy for the neutral pions and increase for the $\eta$-mesons (this confirms the result which is mentioned in the beginning of the capture). The last result can also meant that: the density of the regions increase for the $\pi^0$ – mesons[23] and decrease for the $\eta$-mesons. In these interval of energies, the ratios of lengths for the eta mesons to the ones for the neutral mesons (see Table 5) are:

$$\frac{L_I^{42}}{L_I^{41}} = 2.2 \pm 0.3 \; ; \frac{L_I^{52}}{L_I^{51}} = 3.2 \pm 0.2 \; ; \frac{L_{II}^{52}}{L_{II}^{51}} = 5.2 \pm 0.6 \quad (1)$$

The ratio of the eta mesons mass to one of neutral mesons is $\frac{m_\eta}{m_{\pi^0}} = \frac{0.547 \text{ GeV}/c^2}{0.135 \text{ GeV}/c^2} \cong 4$ and it is compatible with the values of ratios from the equations (1) and one can write:

$$<L_\eta>:<L_{\pi^0}> \cong m_\eta : m_{\pi^0},$$

---

[23] The equations: $b_I^{51} \cong 2b_{II}^{51} \cong 4b_{III}^{51} \cong 8b_{IV}^{51}$ ; $b_i^{51} \cong \frac{1}{2^{(K-1)}} b_I^{51}$, $K=2\div4$ are observed.

here $<L_\eta>$ and $<L_{\pi^o}>$ are some averaged values of the lengths for the eta meson and neutral pion respectively. This is a very important result in favor of parton string fragmentation dynamics. Because in string theory the masses of elementary particles and their energy are defined by the intensity of string vibration and strangeness of the string stretch. The lengths of the regions $L_K^c$ have to be directly proportional to string vibration and strangeness of the string stretch. To produce massive particles more intensive vibrations (or more strangeness of the string stretch) are necessary. That is why the lengths for the eta mesons greater than one for the neutral mesons. The increasing the lengths for the eta mesons' $p_T$ regions accompanied with increasing of the values of $b_k^c$. It can be meant that the $\eta$-mesons were produced at the less values of $\alpha_S$ compared with one for neutral mesons (because as we have mentioned in the Introduction the values of $b_K^c$ are connected with the values of the $\alpha_S \cong \left[ ln\left(\frac{q^2}{\Lambda^2}\right)\right]^{-1}$, $Q^2 = -q^2 \cong \frac{1}{b_K^{c^2}}$ ).    STOP

Table 5. The values of the ratios for $L_K^c$, $a_K^c$ and $b_K^c$ in the regions in the events with $c_1$ to the ones in the events with $c_2$ (in the 3 regions :I,II ,II)

| $c_2$:$c_1$ | | 31:3 | 11:1 | 12:11 | 41:4 | 42:41 | 51:5 | 52:51 |
|---|---|---|---|---|---|---|---|---|
| $L_K^c$ | I | 0.9±0.1 | 1.0±0.1 | 1.2±0.2 | 0.65±0.05 | 2.2±0.3 | 1.00±0.04 | 3.2±0.2 |
| | II | - | 0.7±0.1 | 0.9±0.1 | 0.33±0.05 | - | 0.55±0.02 | 5.2±0.6 |
| | III | - | - | - | - | - | - | - |
| $a_K^c$ | I | 0.3±0.1 | 0.31±0.08 | 0.012±0.009 | 0.7±0.2 | 0.04±0.01 | 1.0±0.2 | (8.2±2.3)10$^{-3}$ |
| | II | - | 0.2±0.1 | - | 8.3±3.8 | - | 0.3±0.1 | (4.5±2.0)10$^{-4}$ |
| | III | - | - | - | - | - | - | - |
| $b_K^c$ | I | 1.03±0.08 | 0.97±0.07 | 0.8±0.1 | 1.22±0.06 | 0.64±0.04 | 1.23±0.07 | 0.45±0.03 |
| | II | - | 1.0±0.1 | 1.0±0.2 | 1.9±0.1 | - | 1.08±0.07 | 0.29±0.02 |
| | III | - | - | - | - | - | - | - |

The Table 6 shows the ratios for the values of $L_K^c$ ($L_{II}^c: L_I^c$, $L_{III}^c: L_{II}^c$, $L_{IV}^c: L_{III}^c$) and inverse ratios for the free fitting parameters $b_K^c$ (($b_{II}^c: b_I^c)^{-1}$, $(b_{III}^c: b_{II}^c)^{-1}$, $(b_{IV}^c: b_{III}^c)^{-1}$) for the different events (the index $c$). The values of the $p_T$ at the boundaries ($J_{I-II}^c$, $J_{II-III}^c$, $J_{III-IV}^c$) of the $p_T$ regions are also included (rows 5,8 and 11). First we would like to note that the boundary values for the I-II $p_T$ regions don't depend on energies of the collisions and characteristics of the particles produced in the events with $c$=3,1,11 and $c$=4 ($J_{I-II}^c \cong$ 4 GeV/c, first group of regions) but for the second group of regions ($c$=41, 5,51), the values of the $J_{I-II}^c \cong$ 2.2 ÷ 2.6 GeV/c and has anomaly high values: 7 GeV/c for the regions in the events with $c$=52 (the II group of events). The boundary values for the *II-III* $p_T$ regions are also classified into further two groups, for the first group of regions (in the events with c=3,1,11 and 4 ) the values of $J_{II-III}^c \cong$ 15÷20 GeV/c and it is around 5-7 GeV/c for the second group of regions in the events with c=41, 5,51.

Table 6

| rato / c | $L_V^c: L_{IV}^c$ | $L_{IV}^c: L_{III}^c$ | $(b_{IV}^c: b_{III}^c)^{-1}$ | $J_{III-IV}^c$ | $L_{III}^c: L_{II}^c$ | $(b_{III}^c: b_{II}^c)^{-1}$ | $J_{II-III}^c$ | Note | $L_{II}^c: L_I^c$ | $(b_{II}^c: b_I^c)^{-1}$ | $J_{I-II}^c$ | Note |
|---|---|---|---|---|---|---|---|---|---|---|---|---|
| 3 | | - | - | - | - | - | 13.2±1.4 | | 3.3±0.4 | 3.4±0.2 | 4.0±0.3 | |
| 1 | | - | - | - | 4.4±1.0 | 5.6±0.6 | 21.6±2.5 | | 5.0±0.9 | 4.2±0.4 | 3.6±0.4 | $L_{II}^c: L_I^c \cong (b_{II}^c: b_I^c)^{-1}$ |
| 11 | | - | - | - | - | - | 15.9±0.8 | | 3.6±0.3 | 3.0±0.4 | 3.5±0.2 | |
| 4 | | - | - | - | 6.4±0.9 | 5.2±0.4 | 17.2±1.4 | $L_{II}^c: L_I^c \cong (b_{II}^c: b_I^c)$ | 3.3±0.4 | 3.6±0.2 | 4.0±0.3 | |
| 41 | | - | - | - | - | - | 7.0±0.3 | | 1.7±0.2 | 2.2±0.1 | 2.60±0.07 | |
| 5 | | - | - | - | - | - | 6.94±0.04 | | 2.15±0.05 | 1.89±0.08 | 2.20±0.04 | $L_{II}^c: L_I^c \ne (b_{II}^c: b_I^c)^{-1}$ |
| 51 | 22.4±1.2 | 3.1±0.4 | 2.0±0.2 | 9.1±0.3 | 1.7±0.1 | 1.9±0.2 | 4.80±0.07 | | 1.18±0.06 | 2.2±0.2 | 2.20±0.07 | |
| 52 | | - | | - | - | - | - | | 1.9±0.3 | 3.4±0.2 | 7.0±0.4 | |

The ratios $L_{II}^c: L_I^c$ also show of the existing two groups of regions (see Table 6):

$$L_{II}^3: L_I^3 \cong L_{II}^1: L_I^1 \cong L_{II}^{11}: L_I^{11} \cong L_{II}^4: L_I^4 \cong 3 \div 5 \text{ , the first group,}$$

$$L_{II}^{41}: L_I^{41} \cong L_{II}^5: L_I^5 \cong L_{II}^{51}: L_I^{51} \cong L_{II}^{52}: L_I^{52} \cong 1 \div 2 \quad \text{the second group.}$$

The ratios $L_{III}^c: L_{II}^c$ point to existing of two groups of regions too:

$$L_{III}^1: L_{II}^1 \cong L_{III}^4: L_{II}^4 \cong 5 \text{ (first group),}$$

$$L_{III}^{51}: L_{II}^{51} \cong 2 \; ; \; L_{IV}^{51}: L_{III}^{51} \cong 1 \div 2 \quad \text{(second group)}$$

In terms of parton string fragmentation, to discuss the results for these ratios, one can say that the particles from the I group of regions are produced through the strings breaking up mainly to 3-5 strings/partons, though particles of II group regions-are formed through the strings breaking up mainly to 2 strings/partons. As mentioned above that the *I* group regions are mainly the ones which are observed at energies 0.9 and 2.76 TeV and second group of regions are observed at energies 7 and 8 TeV. Then the values of the parameters $L_K^c$ and $b_K^c$ (at linearly increasing $a_K^c$) for the first group of regions don't depend on colliding energy and characteristics of the produced particles, though for the *II* group of regions the values of parameters $L_K^c, b_K^c$ and $a_K^c$ depend strongly on the energy and the characteristics of the produced particles. That could be a hint that there are two types of strings responsible for the observed groups.

One more observation from the Table 6 is that for the first group of regions (in the events with c=3,1,11 and c=4) $L_{II}^c: L_I^c \cong (b_{II}^c: b_I^c)^{-1}$ but for the second group of regions (in the event with c=41,5,51 and 52) $L_{II}^c: L_I^c \neq (b_{II}^c: b_I^c)^{-1}$.

The values for the ratios of free fitting parameters $a_K^c$ (Table 7) also show the existing same two groups.

Table 7

| Ratios<br>Event (c) | $a_{IV}^c: a_{III}^c$ | $a_{III}^c: a_{II}^c$ | $a_{II}^c: a_I^c$ |
|---|---|---|---|
| 3 (ch. pp 0.9 TeV) | - | - | $(3.8 \pm 1.5)10^{-4}$ |
| 1 (ch. pp 2.76 TeV) | - | $(1.3 \pm 1.0)10^{-2}$ | $(5.8 \pm 2.8)10^{-4}$ |
| 11 ($\pi^0$ pp 2.76 TeV) | - | - | $(4.1 \pm 1.7)10^{-4}$ |
| 4 (ch. pp 7.0 TeV) | - | $(1.0 \pm 0.4)10^{-3}$ | $(8.9 \pm 5.0)10^{-5}$ |
| 41 ($\pi^0$ pp 7.0 TeV) | - | - | $(1.2 \pm 0.3)10^{-2}$ |
| 5 (ch. pp 8.0 TeV) | - | - | $(5.3 \pm 1.0)10^{-2}$ |
| 51 ($\pi^0$ pp 8.0 TeV) | $(1.5 \pm 0.8)10^{-1}$ | $(2.2 \pm 1.2)10^{-3}$ | $(1.7 \pm 0.6)10^{-2}$ |
| 52 ($\eta$ pp 8.0 TeV) | - | - | $(9.5 \pm 3.7)10^{-4}$ |

## 7. Conclusion.

The study of the inclusive spectra of the charged particles, neutral pions and eta mesons produced in the *pp* collisions at LHC energies shows several $p_T$ regions, which are characterized by the length of the regions $L_K^c$ and two free fitting parameters $a_K^c$ and $b_K^c$. We observed two region groups: the first one contains mainly the charged particles at energies 0.9-7.0 TeV and pions produced at energies 0.9 and 2.76 TeV; the second group contains the pions and eta mesons at energies 7-8 TeV. The $L_K^c$ and $b_K^c$ values for the first group don't depend on colliding energy and the characteristics of the produced particles, whereas for the second region they show strong dependencies. The length values of the $p_T$ regions' for the neutral pions are less than for the eta mesons. The ratios of the averaged lengths for the eta mesons to neutral mesons is directly proportional to the ratio of their masses: $<L_\eta>: <L_{\pi^0}> \cong m_\eta: m_{\pi^0}$. This result could be evidence of parton string fragmentation dynamics, because in string theory the masses of elementary particles and their energies are defined by the intensity of string vibration, and we propose that the values of $L_K^c$ must be proportional to string vibration and strangeness of the string stretch. The increase in lengths for the eta mesons' region is accompanied with increasing of the values of the $b_k^c$. It can mean that the $\eta$-mesons were produced at less values of $\alpha_S$ compared to neutral mesons due to production of the strange quarks in case of eta mesons. We propose that considering string fragmentation and hadronization dynamics, the particles from the group *I* are produced through string decays mainly into 3-5 strings, whereas group *II* particles are mainly produced through the strings decay into 2 strings. We claim that the observed two region groups are formed under the influence of the collision energy (defined by the tension of the original strings), mass of the produced particles (determined by the vibration of the original strings) and the charge and quantum number of the original strings. The existence of the 2 groups of regions could indicate that there are two types of strings responsible for the observed division.

In the end I would like to acknowledge the COMSATS University Islamabad, which provided suitable platform and all possible facilities to perform the analysis, Aziza Suleymanzade, Dr. Kamal H. Khan and Dr. Ali Zaman for their essential help during preparing the text.